\documentclass{iau} 
\usepackage{graphicx}

\def\la{\mathrel{\mathchoice {\vcenter{\offinterlineskip\halign{\hfil
$\displaystyle##$\hfil\cr<\cr\sim\cr}}}
{\vcenter{\offinterlineskip\halign{\hfil$\textstyle##$\hfil\cr
<\cr\sim\cr}}}
{\vcenter{\offinterlineskip\halign{\hfil$\scriptstyle##$\hfil\cr
<\cr\sim\cr}}}
{\vcenter{\offinterlineskip\halign{\hfil$\scriptscriptstyle##$\hfil\cr
<\cr\sim\cr}}}}}
\def\ga{\mathrel{\mathchoice {\vcenter{\offinterlineskip\halign{\hfil
$\displaystyle##$\hfil\cr>\cr\sim\cr}}}
{\vcenter{\offinterlineskip\halign{\hfil$\textstyle##$\hfil\cr
>\cr\sim\cr}}}
{\vcenter{\offinterlineskip\halign{\hfil$\scriptstyle##$\hfil\cr
>\cr\sim\cr}}}
{\vcenter{\offinterlineskip\halign{\hfil$\scriptscriptstyle##$\hfil\cr
>\cr\sim\cr}}}}}

\title[IAU 361.~~Lithium in red supergiants] 
{Strong lithium lines in red supergiants at different metallicities}

\author[I. Negueruela et al.]   
{Ignacio Negueruela$^1$, Javier Alonso-Santiago$^2$, Ricardo Dorda$^3$ \and Lee. R. Patrick$^1$}

\affiliation{$^1$Departamento de F\'{\i}sica Aplicada, Facultad de Ciencias, Universidad de Alicante, Carretera de San Vicente s/n, E03690, San Vicente del Raspeig, Spain \\ email: {\tt ignacio.negueruela@ua.es}
\\[\affilskip]
$^2$ INAF--Osservatorio Astrofisico di Catania, via S. Sofia 78, 95123 Catania, Italy
\\[\affilskip]
$^3$ Instituto de Astrof\'{\i}sica de Canarias,  V\'{\i}a L\'actea s/n, E38200, La Laguna, Tenerife, Spain}

\pubyear{2022}
\volume{361}  
\jname{Title of your IAU Symposium}
\editors{A.C. Editor, B.D. Editor \& C.E. Editor, eds.}
\begin{document}

\maketitle

\begin{abstract}
Current models of stellar evolution predict that stars more massive than $\sim6\:$M$_{\odot}$ should have completely depleted all lithium (Li) in their atmospheres by the time when they reach the He core burning phase. Against this, a non-negligible number of red giants with masses $\ga6\:$M$_{\odot}$ presenting strong Li lines have recently been reported. Motivated by this finding, we have carried out a spectroscopic survey of red supergiants (RSGs) in the Perseus Arm and a selection of young open clusters in the Magellanic Clouds to assess the presence of the Li\,{\sc i}~6708\AA\ doublet line. Based on a sample of $>70$ objects, close to one third of RSGs in the Perseus Arm display noticeable Li lines, with perhaps a trend towards a lower fraction among more luminous stars. The samples in the Magellanic Clouds are not as large, but hint at a metallicity dependence. Twenty one RSGs in 5 LMC clusters show a very high fraction of Li detection, around 40\%. Conversely, 17 RSGs in 5 SMC clusters give only one secure detection. The interpretation of these observational results is not straightforward, but a mechanism for Li production seems most likely. Further characterisation work is ongoing, while theoretical studies into this matter are urgently needed.
\end{abstract}

\firstsection 
              
\section{Introduction}

Lithium (Li) is the heaviest element created during primordial nucleosynthesis. In the Milky Way, low-metallicity stars consistently display Li abundances around $A (\mathrm{Li})=2.2$ (Spite plateau), while stars with metallicity ($Z$) more similar to the Sun's show higher values, around $A (\mathrm{Li})=3.2$ (e.g. \cite[Bensby \& Lind 2018]{bensby18}). The lower value has been interpreted as the primordial abundance -- although it is very significantly lower than predictions based on models of Big Bang nucleosynthesis --, while the higher value in younger stars suggests the existence of a production mechanism through the aeons (\cite[Randich \& Magrini 2021]{rama21}, and references therein).  

Lithium is easily destroyed at temperatures slightly above $T_{\mathrm{eff}}=2\times10^6\:$K. It captures a proton and disintegrates into two He nuclei. In stars with convective envelopes, Li is transported from the outer layers to higher temperature zones and depleted on a timescale of a few to tens of Ma (e.g. \cite[Jeffries 2006]{jeffries2006}). On the other hand, intermediate-mass stars with radiative envelopes can keep their original Li abundance in a thin outer layer. As the star leaves the main sequence and starts to swell, Li is diluted. Later, during the first dredge-up, strong Li depletion further reduces its abundance. This evolution is nicely illustrated in clusters containing stars with masses $\la 2\:$M$_{\odot}$ (e.g. \cite[Twarog et al.\ 2020]{twarog20}), where Li is severely depleted in subgiants and falls below detectability in most red giants, including clump stars (e.g.\ \cite[Anthony-Twarog et al.\ 2021]{antwarog21}).  More massive stars are expected to behave similarly, although Li lines are not seen in stars with $T_{{\rm eff}} > 8\,500\:$K (\cite[Lyubimkov 2016]{lyubimkov16}). Studies of low-luminosity supergiants \cite{luck77} and stars in moderately young clusters \cite{luck94} did not hint at significant differences with respect to less massive stars with radiative envelopes. 

Lithium is not only destroyed in stars. Under some circumstances, it can also be created. A small fraction ($\sim1$\%) of K-type giants show high Li abundances, in some cases well above those observed in main-sequence stars, strongly hinting at a production mechanism. Different authors have argued for a transient phenomenon, connected to the star's evolution (e.g. \cite[de la Reza~2020]{reza20}). Recent studies based on large samples of field stars suggest that this phenomenology may be related to the He flash (\cite[Zhang et al.\ 
2021]{zhang21}). Many Li-rich stars seem to be in the clump, and several authors have suggested that more than one mechanism must be at play (e.g.\ \cite[Martell et al.\ 2021]{martell21}; \cite[Zhou et al.\ 2022]{zhou22}). High Li abundances are also seen in    massive AGB stars. These are very luminous giants of late-M type, believed to represent the last phase in the life of stars with initial masses $4$\,--\,$9\:$M$_{\odot}$. Evolved AGB stars, understood to be close to the end of the thermal-pulse phase, show high Rb abundances and a very wide range of $A (\mathrm{Li})$. In the Magellanic Clouds, such objects display $-7\la M_{\mathrm{bol}}\la-6$ \cite{g-h07}. Other similar objects, assumed to be in a slightly earlier phase, have been found to display very high $A (\mathrm{Li})>3$, while not showing high abundances of s-process elements \cite{g-h13}. This is taken as support for models that predict strong Li production at the start of the thermal-pulse phase, when Hot Bottom Burning is activated, as the base of the envelope of the star reaches $T \ga 4\times10^{7}\:$K, and the Cameron-Fowler mechanism \cite{cf71} can proceed  (e.g. \cite{vanraai12}).  In the Cameron-Fowler mechanism, enhanced production of beryllium (${}^7$Be) will occur at the base of the envelope. Convection during the third dredge-up will move it to regions of lower $T_{{\rm eff}}$ in the outer layers, where it may decay into lithium \cite{mazzitelli}.

The behaviour of Li in massive stars has not been studied in depth. \cite{lyubimkov12} observed a sample of Galactic F and G-type supergiants, determining stellar parameters and Li abundances. They found that stars with initial masses $\la 6\:$M$_{\odot}$ may show a wide range of Li abundances, going from non-detectable to the same value as main-sequence stars, a spread that they interpret as a consequence of different initial rotations -- suggesting that stars with very low rotation may keep most of their original surface Li until the first dredge-up, and a reduced abundance $A$(Li)$\la 1.3$ after it has happened. Conversely, \cite{lyubimkov12} barely detect any lithium in any star with estimated mass $\ga 6\:$M$_{\odot}$, a fact that they explain resorting to stellar models that include the effect of rotation (e.g. \cite{brott11}; \cite{ekstrom12}). These models predict that, even for relatively low initial rotational velocities, Li is depleted below detectability by the end of the main sequence, due to rotational mixing. In models without rotation, a precipitous drop in Li abundances happens during the B-type giant phase for stars of 7 to~$15\:$M$_{\odot}$. Such models are also supported by limited observations of the behaviour of Be, another light element \cite{proffitt16}. Nevertheless, recent observations of red supergiants in open clusters, with estimated masses 7\,--\,$9\:$M$_{\odot}$, have shown a non-negligible fraction of objects with strong Li lines \cite{negueruela20}, in open contradiction with models and previous results. In an attempt to check if this behaviour extends to higher masses, we have performed two observational campaigns targeting large populations of RSGs at different metallicities.

\section{Observations}

We conducted an exploratory survey with the 80~cm Telescopi Joan Or\'o at the Montsec Observatory, which operates in robotic mode \cite{vilardell13}. We used the ARES spectrograph with the Red VPH, which provides a resolving power $R\approx10\,000$ over the 630 to 673 nm range. Observations, taken between December 2019 and June 2020, targeted a small sample of RSGs from the Perseus arm, selected from the lists of \cite{dorda18} and the references included there. Since these spectra indicated that a sizable fraction of the RSGs present Li lines, we proceeded to observe large samples of stars.

A sample of Perseus arm RSGs was observed with the high-resolution FIbre-fed Echelle Spectrograph (FIES) attached to the 2.56~m Nordic Optical Telescope (NOT; La Palma, Spain) during a run on 2020, October 2\,--\,5. FIES is a cross-dispersed high-resolution echelle spectrograph that covers the 370\,--\,830~nm range. We used the large aperture, which provides a resolving power $R\approx25\,000$. The spectra were homogeneously reduced using the FIEStool software in advanced mode. We observed $>50$ RSGs, distributed over the Northern sky, most of them from the Cassiopeia region of the Perseus Arm. When combined with over 20 RSGs from the region of Per~OB1 in \cite{deburgos20}, observed with the same instrument, this results in a sample of over $70$ Perseus Arm RSGs.

Samples of RSGs at different metallicities were obtained by observing RSGs in Magellanic Cloud clusters. These objects were observed with \textit{X-shooter}, mounted on the VLT UT3 (Melipal). Most of the observations used the UVB arm with an $1.3^{\prime\prime}$ slit ($R\approx4\,000$) and the VIS arm with a $0.9^{\prime\prime}$ slit ($R\approx9\,000$). Observations were obtained in service mode between 2020 October 13 and December 29. The targets observed were selected to belong to open clusters of different ages, in order to cover a wide range of masses. The clusters observed are listed in Table~\ref{clusters}.

\begin{table}
  \begin{center}
  \caption{Clusters containing RSGs observed with \textit{X-shooter}. An estimate of the cluster age and the number of RSGs observed in each cluster are given.}
  \label{clusters}
 {\scriptsize

  \begin{tabular}{l|c|c||l|c|c}
   \noalign{\smallskip}
  \hline 
   \noalign{\smallskip}
  \multicolumn{3}{c||}{SMC} & \multicolumn{3}{|c}{LMC}\\
  \hline
Name & Age (Ma) & $N$ & Name & Age (Ma) & $N$\\
\hline
NGC~299 & 20 & ~3~ &  NGC~2100 & 18 &  ~7~ \\
NGC~376 & 25 & ~4~ &  NGC 1772 & 30 & ~4~ \\
Bruck 71 & 35 & ~3~ &  NGC~1818 & 30 & ~2~ \\
NGC~241/2 & 60 & ~5~ & NGC~1711 & 40 & ~4~ \\
NGC~256 & 90 & ~2~ & NGC~1755 & 60 & ~4~ \\
\hline
  \end{tabular}
  }
 \end{center}

\end{table}

\section{Results}

Analysis of the spectra obtained is ongoing, in order to provide a full characterisation of all the target stars that will serve as a context for the detection of Li lines. Meanwhile, visual inspection can inform us about the basic properties of the stars.

Among the Perseus Arm sample, we find a high fraction of stars with Li lines. About 25\% of the stars of all spectral types (stars observed fall mainly in the K5\,--\,M4 range, typical of Milky Way RSGs -- see \cite[Dorda et al.\ 2018]{dorda18}) display strong Li lines. A few others have weak Li lines that appear blended with neighbouring weak metallic features. This result is in agreement with a recent report by \cite{fanelli22}, based on a smaller sample. 

Our RSG sample in the Perseus Arm consists mostly of field stars, which cannot be assigned ages, and thus masses, to compare to evolutionary tracks. To address this issue, all our targets in the Magellanic Clouds are members of open clusters, which can provide an evolutionary context for them. In Table~\ref{clusters}, we list the clusters targeted, together with an estimate of their ages (based on isochrones without binary interaction and the assumption that the \textit{bulk} of the RSGs determines the age). 

\begin{figure}[b]
\begin{center}
\includegraphics[width=10.5cm, angle=-90, clip]{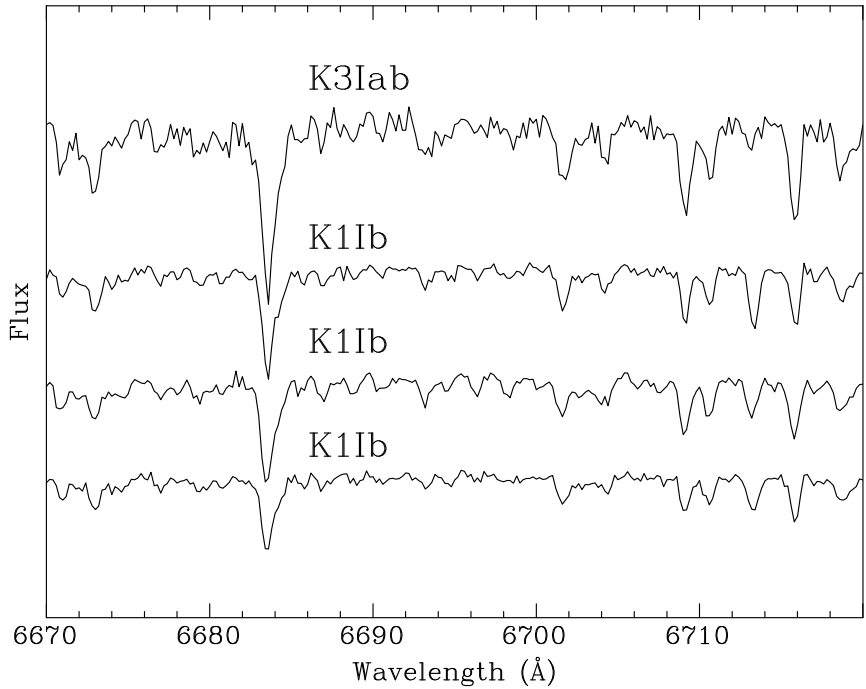} 
 \caption{A small section of the \textit{X-shooter} spectra of 4 RSGs in the LMC cluster NGC~1711. The Li\,{\sc i}~6708\AA\ line, displaced to close to 6714\AA\ because of the systemic velocity of the LMC, is seen in the three bottom stars.}
   \label{spect}
\end{center}
\end{figure}

Our Magellanic samples are smaller than that from the Milky Way, but likely large enough to be statistically representative. The LMC sample includes one of the most well-populated RSG clusters, NGC~2100, which is likely younger than 20~Ma, and a range of clusters extending to the well studied NGC~1755, which is believed to contain RSGs of $\sim 7\:\mathrm{M}_{\odot}$. In total, we have observed 21 RSGs. Stars with Li lines are found in all clusters, except in NGC~1818, where only 3 could be observed and one turns out to have an early (G) spectral type. As an example, Fig.~\ref{spect} shows a small section of the spectrum of the 4 RSGs observed in NGC~1711, three of which display a clearly noticeable Li line. In total, about 40\% of the stars observed have lithium. Given the small sample size, this value in consistent with that found in the Milky Way. 

In the SMC, the number of large young open clusters is much smaller than in the LMC, and we could not find many suitable targets. As a consequence, our clusters cover a wider range in ages, reaching NGC~256, where the RSGs are expected to have $\sim 6\:\mathrm{M}_{\odot}$. Among 17 stars observed, there is only one with a moderately strong Li line, in Bruck~71. Two stars in NGC~241/2 may have weak Li present. Although the sample is small, this result very strongly suggests a different behaviour from that seen at higher $Z$.

\section{Discussion}

The most massive intermediate-mass stars ($M_{*}\ga 6\:\mathrm{M}_{\odot}$) and most high-mass stars (except for the most massive ones) appear as RSGs (i.e. are classified as luminosity class I objects) when evolved. Thus, RSGs below a threshold mass -- that likely depends on $Z$ and probably also the evolutionary history of the star --  will behave as intermediate-mass stars, entering the AGB and leaving behind a white dwarf as final remnant (see, e.g. discussion in \cite{neg17}). Modern models (e.g. \cite{poelarends08}; \cite{doherty15}) indicate quite high values (approaching $10\:$M$_{\odot}$) for this transition at $Z_{\odot}$, although binary interaction may have a significant effect on the final outcome. At $Z_{\mathrm{SMC}}$, the minimum mass to explode as a supernova should be closer to  $8\:$M$_{\odot}$. The reliability of all these models, however, depends on our ability to understand the complex physics governing the transport of mass and angular momentum between the core and the outer layers and the behaviour of convection at different boundary layers.

Given our current understanding, Li should not be present in the atmospheres of evolved massive stars -- especially if they have started their lives as moderate or fast rotators. Despite this, we find strong Li lines -- much stronger than seen in any normal red giant -- in a substantial fraction of RSGs in the Milky Way and the LMC. How can these observations be interpreted?

At present, there is no obvious explanation for these results. A number of scenarios can be invoked, but they all suffer from severe shortcomings. Among these scenarios, we can list:

\smallskip

\begin{itemize}
    \item \textbf{The lithium observed is preserved throughout the life of the star.} This would require that no rotational mixing and no mass loss has occurred. Given that only a fraction of the RSGs have Li and this fraction is smaller in the SMC, it would be tempting to identify the RSGs with Li with the original slow rotators. However, it must be noted that this preservation is not seen in lower mass stars (which generally rotate slowly). Moreover, it is difficult to envisage how stars of $>10\:\mathrm{M}_{\odot}$ may have reached the RSG phase without substantial mass loss.
    \item \textbf{RSGs with Li are in fact early-AGB stars producing Li via the Cameron-Fowler mechanism}. In recent years, the use of isochrones based on tracks including binary interaction has resulted in older age determinations for young open clusters, under the assumption that stars close to the main-sequence turn-off are mostly mass gainers (and, hence, blue stragglers). We could thus assume that most of the RSGs that we have observed, even those in Per~OB1, are less massive than $\la12\:$M$_{\odot}$, and behave as AGB stars. Although this interpretation can provide us with an already-tested physical basis for the presence of Li, it does not hold on evolutionary terms. A massive star spends $\sim10$\% of its lifetime as a He-core burning RSG, but an intermediate-mass star spends $<1$\% of its lifetime as an E-AGB star. There are simply too many RSGs with Li (and specifically, too many RSGs with Li in open clusters) for this interpretation to be correct.
    \item \textbf{The RSGs with Li have engulfed brown dwarf companions.} Engulfment of giant planets has been proposed as an explanation for the existence of some Li-rich giants \cite{aguilera16}. For a massive star, a very hefty planet would be needed to provide enough Li to be detected, and brown dwarfs are likely necessary. The viability of such a process has not been studied, but the much higher fraction of RSGs with Li as compared to solar-mass giants renders this scenario suspect.
    \item \textbf{A mechanism to take Li to the surface works in RSGs}. The Cameron-Fowler mechanism requires third dredge-up convection to penetrate below the H-shell burning layer and retrieve newly created Be to the surface, where it can decay to lithium. A mechanism for penetrating the active H layer in RSGs is not obvious, but it might provide an explanation to the observed high frequency. The lower fraction at $Z_{\mathrm{SMC}}$  will serve as a consistency check on any mechanism devised. 
\end{itemize}

\smallskip

In summary, the detection of Li in a high fraction of RSGs is extremely surprising under the light of our current stellar models. Further theoretical developments are certainly needed to provide an understanding of this phenomenon.

\begin{acknowledgements}

Partially based on observations collected at the European Organisation for Astronomical Research in the Southern Hemisphere under ESO programme 106.212J.001 and on observations made with the Nordic Optical Telescope, operated by the Nordic Optical Telescope Scientific Association at the Observatorio del Roque de los Muchachos (La Palma, Spain)
of the Instituto de Astrof\'{\i}sica de Canarias. We thank the director of the Montsec Observatory for DDT time on the TJO.
This research is partially supported by the Spanish Government under grants PGC2018-093741-B-C21/C22 (MICIU/AEI/FEDER, UE), and by the Generalitat Valenciana under grant PROMETEO/2019/041.
\end{acknowledgements}

\bibliographystyle{aa}

\end{document}